\documentclass[twoside,reqno]{HERON}
\usepackage{epsfig,cite,colordvi}
\usepackage{graphicx}
\usepackage{url}
\newcommand{\cd}{\makebox[0.08cm]{$\cdot$}}
\usepackage{amssymb,amsmath,amscd,epsf}
\usepackage{times}
\usepackage{makeidx}
\begin{document}

\title{Systems Dominated by Exchange Particles}

\runningheads{Systems Dominated by Exchange Particles}{V.A. Karmanov}

\begin{start}

\author{V.A. Karmanov}{1}

\index{Karmanov, V.A.}

\address{Lebedev Physical Institute of Russian Academy of Sciences,\\ Leninsky prospect 53, 119991 Moscow, Russia}{1}


\begin{Abstract}
As well known, the spectrum of a non-relativistic two-body system interacting by the Coulomb potential is the Balmer series $E_n=\frac{\alpha^2m}{4n^2}$
produced by the Schr\"odinger equation. In 1954, Wick and Cutkosky have found, in the Bethe-Salpeter equation framework, that for $\alpha>\frac{\pi}{4}$
the relativistic effects result in new levels (in addition to the Balmer series).  However, the physical nature of these new states remained unclear and therefore their existence was being questioned. We have recently shown that these extra states are dominated by the exchange (massless) particles, moving with speed of light. That's why they did not appear in the non-relativistic (Schr\"odinger) framework. 
\end{Abstract}
\end{start}

\section{Introduction}\label{intro}
Hydrogen atom played outstanding role in establishing the quantum mechanics. Description of its spectrum by the formula 
\begin{equation}\label{balmer}
E_n=-\frac{\alpha^2m}{4n^2}, 
\end{equation}
firstly found empirically by Bohr  and then derived  by Pauli (by the matrix method) and, soon after, independently, by Schr\"odinger (from his equation),  
was one of the first great successes of quantum mechanics.
These derivations have been done in non-relativistic approach. However, it turned out that the relativistic effects, taken into account in 1954 by Wick \cite{Wick} and Cutkosky \cite{Cutkosky} in the framework of the Bethe-Salpeter (BS) equation \cite{bs}, not only shift the position of $E_n$, but also result in appearance of new levels, which are absent in the Schr\"odinger equation with the Coulomb potential and are not described by the Balmer series. The physical nature of these new states  remained unclear during almost 70 years. Some researchers believed  that these states are indeed predicted in the BS framework and therefore must exist in nature, others assumed  that they indicated a defect of the BS equation (see for review \S 8 in \cite{nakanishi}). In any case, prediction in a Coulomb system, among the Balmer series,  of  the extra states is an event of fundamental importance. 

The nature of these extra states was clarified in our recent papers \cite{EPJC,universe}. We calculated for these states, called in literature "abnormal", the percentage of the constituent (valence, massive)  particles (their contribution to the full normalization of the state vector, equaled to 1). In sharp contrast to the normal states, this contribution turned out to be less than 10\% or, depending on the state, even less than 1\%. This means that these states are dominated, at least,  for 90-99\% by the exchange (massless) particles. These unusual systems can be distinguished from the "normal" ones, not only by unusual sequence of their levels, but also by behavior of their elastic and transition electromagnetic form factors. 
These our results will be presented below.

Plan of this article is the following. In Sec. \ref{BS}, we present the results found by Wick and Cutkosky. Namely,  we will show that the BS equation reproduces the Balmer series and, in addition, it has extra abnormal solutions, disappearing in the non-relativistic limit. Contribution of the two-body sector to full normalization of the state vector is presented in Sec. \ref{2body}.  For the normal and abnormal solutions, these contributions are drastically different: of the order of 1 for normal solution and vanishingly small for the abnormal one.
The non-relativistic limit is discussed in Sec. \ref{nonrel}. In Sections \ref{ffs} and \ref{transit} the elastic and transition electromagnetic form factors of normal and abnormal systems are calculated. Sec. \ref{concl} contains the concluding remarks.

\section{Bethe-Salpeter equation and its solutions}\label{BS}
In quantum field theory any system is described by the state vector which we denote as $|p\rangle$, where $p$ is the total four-momentum of the system and $M^2=p^2$ is its mass squared. In non-relativistic domain the state vector is reduced to the wave function $\psi(\vec{r})$. The latter, for any bound state,  can be represented in terms of the Fourier integral:
$$
 \psi(\vec{r})=\int\phi(\vec{k})\exp(i\vec{k}\vec{r})\frac{d^3r}{(2\pi)^3},
$$
\ie,  as superposition of the plane waves $\exp(i\vec{k}\vec{r})$  -- the eigenstates of free Hamiltonian. The relativistic state vector $|p\rangle$ can be represented similarly in any subspace with fixed number of particles $n$. These
subspaces are called the Fock sectors, whereas the corresponding wave functions $\psi_n(\vec{k}_1,\vec{k}_2,\ldots,\vec{k}_n)$ are the Fock components.
On the top of that, since in the relativistic theory  the number of particles is not conserved, $|p\rangle$ is a superposition of the Fock sectors, \ie,
of the states with different number of particles.
The eigenstate equation is an infinite system of equations for the Fock components. In order to solve this system approximately, one usually truncates the Fock decomposition, keeping a few components which are expected as  dominating. Solving system of equations requires, as always, the renormalization, which now is the sector dependent one (the counterterms depend on the Fock sectors). Having found a finite set of the Fock components, \ie, approximate truncated state vector $|p\rangle$,  one can study the physical properties of a system, for example, calculate its electromagnetic form factors. For a review of this approach, see \eg \cite{mstk}.

Another approach to the theory of relativistic bound systems \cite{bs} deals not with the state vector $|p\rangle$ itself and its Fock decomposition, but with the matrix element taken from  the T-product of  the Heisenberg operators $\hat{\varphi}(x)$ between the vacuum state $\langle 0|$ and the state $|p\rangle$, namely:
\begin{equation} \label{bs}
\Phi(x_1,x_2,p)=\langle 0 \left| T\Bigl(\hat{\varphi}(x_1)\hat{\varphi}(x_2)\Bigr)\right|p\rangle\ ,
\end{equation}
$x_{1,2}$ are the four-vectors.
The matrix element $\Phi(x_1,x_2,p)$ is the  BS  amplitude in the 4D coordinate space. Sometimes, the amplitude defined by Eq. (\ref{bs}) is called ``the two-body BS amplitude''. To avoid misunderstandings, we would like to emphasize that this is, to some degree, slang reflecting the fact that this BS amplitude depends on two variables $x_1,x_2$. The state vector  $|p\rangle$ in the definition \mbox{(\ref{bs})} contains all of the Fock components, including the many-body ones. Therefore, the BS amplitude (\ref{bs}) implicitly incorporates  information not only about the two-body Fock sector, but also about the higher ones.
After extracting the factor $\exp \left[-ip(x_1+x_2)/2\right]$ the BS amplitude depends on the relative coordinate $x=x_1-x_2$  that in the momentum space corresponds to dependence on the relative four-momentum $k$: $\Phi=\Phi(k,p)$.
 The latter amplitude satisfies the 
 BS equation:
\begin{equation}\label{bseq}
\Phi(k,p)=\frac{i^2\int iK(k,k',p)\Phi(k',p)d^4k/(2\pi)^4}
{\left[(\frac{p}{2}+k)^2-m^2+i\epsilon\right]\left[(\frac{p}{2}-k)^2-m^2+i\epsilon\right]}.
\end{equation}

For one-boson exchange, in the spinless case, the kernel $K$ reads
\begin{equation}\label{ladder}
iK(k,k',p)=\frac{i(-ig)^2}{(k-k')^2-\mu^2+i\epsilon}.
\end{equation} 
It does not depend on $p$.  For  massless exchange, one should put in (\ref{ladder}) $\mu=0$.

In non-relativistic limit, the kernel (\ref{ladder}) is reduced to
$$
K(\vec{k},\vec{k'})=\frac{g^2}{(\vec{k}-\vec{k'})^2+\mu^2}.
$$
Its Fourier transform results in the Yukawa potential \mbox{$V(r)=-\frac{\alpha}{r}\exp(-\mu r)$} with $\alpha=g^2/(16\pi m^2)$.  For $\mu=0$ it turns into the Coulomb potential\\ $V(r)=-\frac{\alpha}{r}$ entering the Schr\"odinger equation
\begin{equation}\label{schr}
\Delta\psi(\vec{r})+2m[E_n-V(r)]\psi(\vec{r})=0.
\end{equation}
The BS equation (\ref{bseq}) with the kernel (\ref{ladder}) provides a relativistic generalization of the Schr\"odinger equation (\ref{schr}) with the Coulomb potential. 

There exists infinite number of  the solutions $\Phi_n(k,p)$ of the equation (\ref{bseq}) with the kernel (\ref{ladder}) with $\mu=0$ which differ by the integer parameter $n=1,2,\ldots$.  For the S-wave, each solution can be represented in the following integral form   \cite{Wick, Cutkosky}:
\begin{equation}\label{Phi}
\Phi_n(k,p)=\sum_{{\nu}=0}^{n-1}\int_{-1}^1dz  
\frac{-i m^{2(n-{\nu})+1}g_{n}^{\nu}(z)}{\left[m^2-\frac{1}{4}M^2 -k^2-p\cd k\,z-\imath\epsilon\right]^{2+n-{\nu}}},       
\end{equation}
For given $n$, the functions $g_{n}^{\nu}(z),\;\nu=0,\ldots,n-1$ satisfy a system of equations. However, the equation determining the function $g_n^0(z)$ is decoupled from other equations. It has the form:
\begin{equation}\label{gndf}
g_n^{0\,\prime\prime}(z)+\frac{2(n-1)z}{(1-z^2)}g_n^{0\,\prime}(z)
-\frac{n(n-1)}{(1-z^2)}g_n^0(z)+\frac{\alpha}{\pi(1-z^2)Q(z)}g_n^0(z)=0,
\end{equation}
with the boundary conditions $g_n^0(\pm 1)=0$.  Here $Q(z)=1-\frac{M_n^2}{4m^2}(1-z^2)$. 
 It determines the spectrum, \ie, the values $M^2_n$. The function $g_n^0(z)$, found from this single equation,  enters in the system of equations for other functions $g_n^{\nu}(z),\; \nu=1,\ldots,n-1$ and, in its turn, determines these functions. 

The number $n$ in (\ref{Phi}) is the counterpart of the principal quantum number appearing in the Schr\"odinger equation (\ref{schr}). Due to the relativistic perturbative correction of the next order in $\alpha$,  the Balmer series obtains the form \cite{FFT} (compare with (\ref{balmer})):
\begin{equation}\label{Enrel}
E_n=-\frac{\alpha^2\left(1+\frac{4}{\pi}\alpha\log\alpha\right)m}{4n^2}.
\end{equation}
However, the relativistic effects are not exhausted by the replacement $\alpha^2 \to \alpha^2\left(1+\frac{4}{\pi}\alpha\log\alpha\right)$
in  (\ref{Enrel}). 
Their most important manifestation is in the fact that for $\alpha> \frac{\pi}{4}$ and for any fixed $n=1,2,\ldots$ the equation (\ref{gndf}) has solutions
for an infinite set of masses $M_{n\kappa}$ and the corresponding eigenfunctions $g^{\nu}_{n\kappa}(z)$ enumerated by the new quantum number 
$\kappa=0,1,2,\ldots$ In other words, for any $n$, there exists an infinite spectrum of excited states.
This is completely analogous to appearance of the infinite spectrum of energies $E_n$ in the Schr\"odinger equation (\ref{schr}). 
These extra excited states are actually highly excited (with energies very close to zero), especially, when $\alpha$ is close to its minimal critical value $\frac{\pi}{4}$.
Namely, the following approximate analytic expression for the abnormal spectrum near the continuum
threshold ($E_{\kappa}=M_{n\kappa}-2m\to 0$) takes place \cite{Wick, Cutkosky}:
\begin{equation} \label{abnspectr}
  E_{\kappa}\simeq -m\exp\left({-\frac{(\kappa-1)\pi}{\sqrt{\frac{\alpha}{\pi}-\frac{1}{4}}}}\right),
\end{equation}
where  $\alpha>\pi/4$ and $\kappa=2,3,\ldots$
At $|E_{\kappa}|/m\ll 1$ this spectrum vs. $\kappa$ does not depend on $n$. For $\alpha\to \frac{\pi}{4}$ all the energies $E_{\kappa}\to 0$.

For $\kappa=0$ and small $\alpha$ the energies $E_n=M_n-2m$
coincide with the non-relativistic ones. If $\alpha\to 0$ ($E_n\to 0$), the ground state function $g^0_{10}(z)$ obtains the following analytical form \cite{Wick}: 
$g^0_{10}(z)=1-|z|$. Substituting it in the integral (\ref{Phi}) for the BS amplitude and then extracting from this amplitude the wave function, we reproduce the well known ground state  hydrogen wave function \cite{universe}. 

The solution $g^0_{n\kappa}(z)$ vs. $z$ has $\kappa$ nodes within the interval $-1<z<1$ and a definite parity \cite{Cutkosky}:
$g^0_{n\kappa}(-z)=(-1)^{\kappa}g^0_{n\kappa}(z)$. The odd solutions do not contribute to the S-matrix \cite{cia,nai}. Therefore, we will  be interested in the solutions with even $\kappa$ only. The binding energies, calculated numerically, for the solutions with $n=1,2$ and $\kappa=0,2,4$ are given in the Table \ref{tab1}.
\begin{table}[!h]
\begin{center}
\caption{Binding energy $|E_{n\kappa}|$ (in the unites of $m$) and two-body norm ($N_2$)
of the low-lying normal  ($n=1,2;\;\kappa=0$) and abnormal
($n=1,2;\;\kappa=2,4$) states, for the coupling
constant $\alpha=5$, for massless exchange $\mu=0$.}\label{tab1}
\smallskip

\small\noindent\tabcolsep=9pt
\begin{tabular}{ccccc}
\hline
\hline
\\[-8pt]
\\[-8pt]
No.&   n & $\kappa$ &  $|E_{n\kappa}|$ & $N_2$ \\
\hline
\\[-8pt]
1&1&0  &   0.999259         &0.65 \\
2&2&0  &   0.208410         & 0.61 \\
3& 1 &2  & $3.51169 \cdot 10^{-3}$ & 0.094 \\
4& 2 &2 & $1.12118 \cdot 10^{-3}$ & 0.077 \\
5&1  &4&$1.54091\cdot 10^{-5}$ & $6.19\cdot 10^{-3}$ \\
6&2  &4& $4.95065    \cdot 10^{-6}$ & $2.06\cdot 10^{-5}$\\
\hline
\end{tabular}
\end{center}
\end{table}

\begin{figure}[h!]
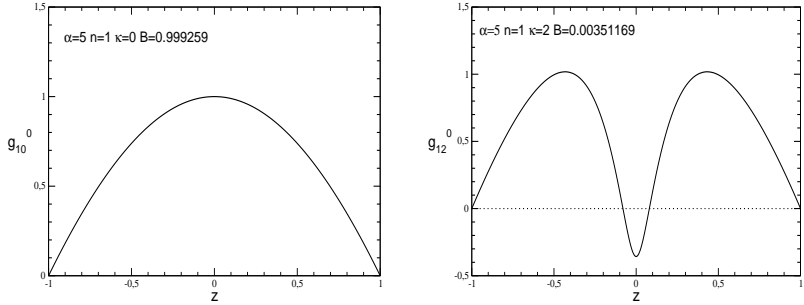

\vspace{0.5cm}
\centering
\epsfxsize=5.cm\epsfysize=4.cm\epsfbox{g100_Nb1.eps}\hspace{0.5cm}
\epsfxsize=5.cm\epsfysize=4cm\epsfbox{g120_Nb3.eps}
\caption{Left panel: $g_{10}^0(z)$ for the  state No. 1 ($n=1$, $\kappa=0$, normal) of the Table \ref{tab1}.\\
Right panel:  $g^0_{12}(z)$ for the  state No. 3 ($n=1$, $\kappa=2$, abnormal) of the Table \ref{tab1}
(adapted from \protect{\cite{EPJC}}).} \label{fig1}
\end{figure}

The solution $g^0_{10}(z)$ (\ie, for $n=1,\kappa=0,\nu=0$), corresponding to No. 1 from the Table \ref{tab1}, is shown in the left panel of Figure
\ref{fig1}.
The solution $g^0_{12}(z)$ (\ie, for $n=1,\kappa=2,\nu=0$), corresponding to No. 3 from the Table \ref{tab1}, is shown in the right panel of Figure
\ref{fig1}. Each solution has $\kappa$ nodes: 0 for $g^0_{10}(z)$ and 2 for $g^0_{12}(z)$.

\section{Weight of two-body sector}\label{2body}

The state vector $|p\rangle$ for the ground state is normalized to 1: $\langle p|p\rangle=\langle 2|2\rangle+\langle 3|3\rangle+\ldots =1$.
The two-body Fock component and its contribution $N_2=\langle 2|2\rangle$ to the full norm equaled to 1 can be expressed via 
the BS amplitude $\Phi(k,p)$ (see \eg,  Sec. 3.3 from \cite{cdkm}). In its turn, BS amplitude $\Phi(k,p)$ is expressed by (\ref{Phi}) through the functions $g_n^{\nu}(z)$. In the end, we obtain expression for $N_2$ in terms of the know functions $g_n^{\nu}(z)$. For example, for the state  with $n=1$, when only one function $g_1^0(z)$, shown for $\kappa=0$ and $\kappa=2$ in Figure \ref{fig1}, contributes in (\ref{Phi}), the expression for $N_2$ reads \cite{EPJC}:
\begin{equation}\label{norm10}
N^{n=1}_2=\frac{1}{384 \pi^2 N_{tot}}\int_{-1}^1\frac{(1-z^2)[g_1^0(z)]^2dz}{[Q(z)]^3},
\end{equation}
$Q(z)$ is the same as in (\ref{gndf}).
The value $N_{tot}$ for any given state  is found from the condition that the elastic electromagnetic form factor at $Q^2=0$ of a system in this state is 1: $F(Q^2=0)=1$. 
We give also the asymptotical formulas \cite{EPJC,universe} for the normal and abnormal states $n=1$, when the absolute value $B$ of the binding energy tends to 0.
For the normal state $n=1$, $\kappa=0$:
\begin{equation}\label{N2a}
N_2(B\to 0)=1+\frac{1}{\pi}\sqrt{\frac{4B}{m}}\log\frac{4B}{m}
\stackrel{B\to 0}{\longrightarrow} 1.
\end{equation}
For the abnormal state $n=1$, $\kappa=2$:
\begin{equation}\label{N2b}
N_2(B\to 0)\propto\sqrt{\frac{B}{m}} \log^2\frac{B}{m}\stackrel{B\to 0}{\longrightarrow} 0.
\end{equation}

The last column of the Table \ref{tab1} shows the contribution of the two-body  sector $N_2=\langle 2|2\rangle$ to the full normalization calculated with the coupling constant $\alpha=5$. We see that it is small ($N_2< 10\%$) for $\kappa=2$ and very small $N_2<1\%$ for $\kappa=4$. This means that 
these state are the many-body ones. Since in this model any sector contains two massive constituents plus massless exchange particles, 
any many-body sector is filled by the exchange particles.
The smallness of $N_2$ and, hence, the dominance of the Fock sectors with $n>2$ means that the abnormal states ($\kappa=2,4,\ldots$) are indeed mainly made of the exchange particles. Strictly speaking, on the ground of $N_2\ll 1$ one cannot exclude that these states are dominated, say, by three-body sector (two massive constituents + one exchange particle), that is $N_3\approx 1$. We will see below that this situation is excluded by behavior of the electromagnetic form factors.


\section{Non-relativistic limit}\label{nonrel}
\begin{figure}[!h]
\centering
\vspace{0.3cm}
\epsfxsize=5.cm\epsfysize=4.cm\epsfbox{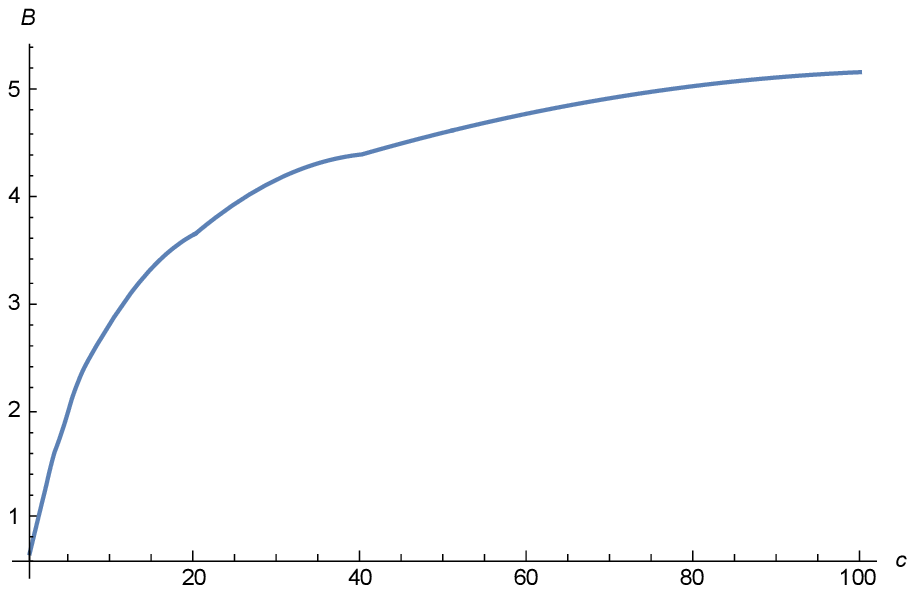}\hspace{0.7cm}
\epsfxsize=5.cm\epsfysize=4cm\epsfbox{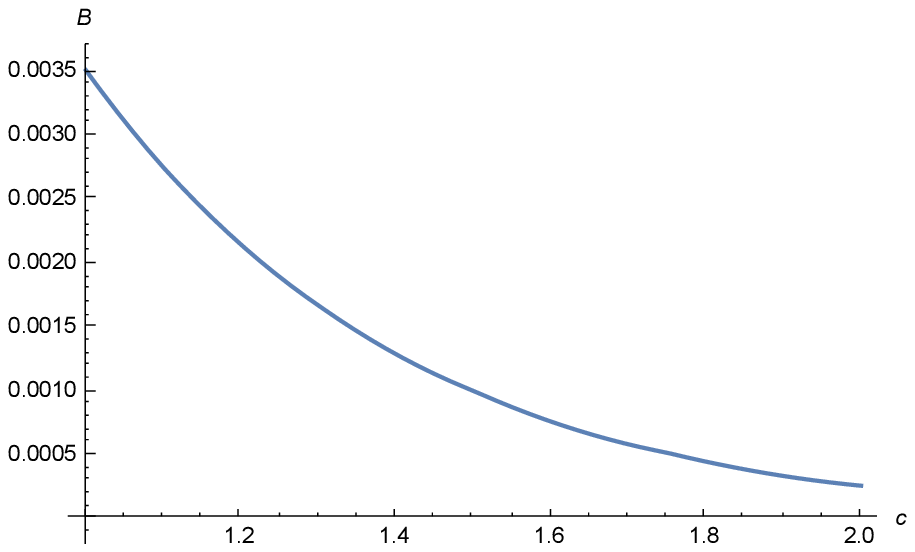}
\caption{Left panel: Dependence of the absolute value of the binding energy $B=|E|$ of the normal state $n=1$, $\kappa=0$ for $\alpha=5$ on speed of light~$c$\\
Right panel: 
The same as in the left panel but for
the abnormal state $n=1$, $\kappa=2$.}
 \label{Bcna}
\end{figure}

The extra states predicted by the BS equation and enumerated by $\kappa=2,4,\ldots$ have relativistic nature and disappear in the non-relativistic limit. It  is instructive to see that explicitly. The principal difference of relativistic physics from the non-relativistic one is existence of the limiting speed of propagation of signal -- the speed of light $c$. If this limiting speed didn't exist, we would be living  in the unique -- non-relativistic -- world, described by the Galilean physics for any speeds and momenta. Therefore, most convenient way to take the non-relativistic limit is to introduce the speed of light $c$ explicitly and take the limit
$c\to\infty$.  The dependence of the total mass $M$ on $c$  is determined by the BS equation. The dependence of the absolute value of the binding energy \mbox{$B=2m-M$} on speed of light $c$ for the normal state $n=1$, $\kappa=0$ (No. 1 in the Table \ref{tab1} for $c=1$) is given in the left panel of Figure \ref{Bcna}. We calculate at first in the units with $c=1$ and then put $c=2,10,20,\dots$.  Figure \ref{Bcna} shows, what happens with the binding energy in smooth transition from relativistic approach to the non-relativistic one. Usually, the relativistic effects are repulsive. Therefore, their weakening when $c$ increases results in increase of attraction and of the the binding energy.
At $c\to\infty$, in the left panel of Figure \ref{Bcna} the binding energy of the normal state found via the BS equation tends to constant (which is given by the Schr\"odinger equation, i.e.,  by  the Balmer series (\ref{balmer})). In the case of abnormal state $n=1$, $\kappa=2$ (No. 3 in the Table \ref{tab1} for $c=1$) this dependence is shown in the right panel of Figure \ref{Bcna}.
 We see that  the binding energy of the abnormal state has opposite behavior: it decreases and tends to zero when $c$ increases. That is, abnormal state disappears in the non-relativistic limit.

\section{Elastic electromagnetic form factors}\label{ffs}
Knowledge of the BS amplitude for given system allows to calculate the electromagnetic form factor of this system.
Expression for the electromagnetic current via BS amplitude has the form:
\begin{eqnarray}\label{ffbs}
J_{\mu}&=&(p_{\mu}+{p'}_{\mu})F(Q^2)
=i\int \frac{d^4k}{(2\pi)^4}(p+p'-2k)_\mu \; (k^2-m^2)
\nonumber\\
&\times&\overline{\Phi}\left(\frac{1}{2}p'-k,p'\right)\Phi\left(\frac{1}{2}p-k,p\right). 
\end{eqnarray}
From here we can express the elastic form factor $F(Q^2)$ via the BS amplitude. This formula can be generalized for the transition
electromagnetic current and form factor (electromagnetic transition between two states in the spectrum). 
Substituting here the BS amplitude in the form (\ref{Phi}), we express form factors through the known functions $g_{n\kappa}^{\nu}(z)$.
We are interested in comparison of the form factors for the normal and abnormal states and transitions between these states.
The details of calculations can be found in \cite{EPJC}. 

\begin{figure}[h!]
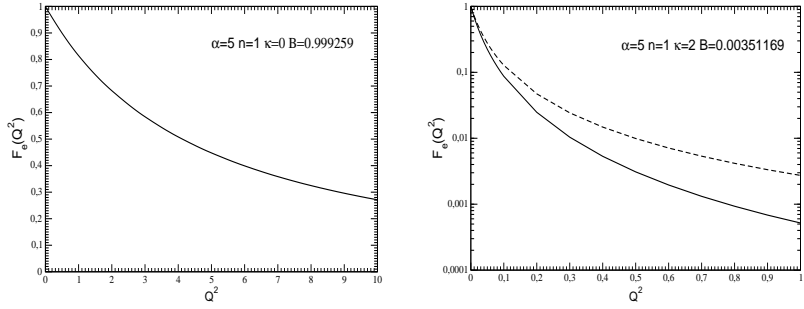

\centering
\vspace{0.3cm}
\epsfxsize=5.cm\epsfysize=4.cm\epsfbox{FQ2_Nb1.eps}\hspace{0.5cm}
\epsfxsize=5.cm\epsfysize=4cm\epsfbox{FQ2_Nb3.eps} 
\caption{Left panel:  the electromagnetic form factor for the normal state with $n=1,\kappa=0$ (No.1 from the Table \ref{tab1}), $\alpha=5$, $B=0.999m$.\\
Right panel: ({\it i}) solid curve:  the electromagnetic form factor for the abnormal state with $n=1,\kappa=2$ (No.3 from the Table \ref{tab1}), $\alpha=5$, $B=0.0035m$; ({\it ii}) dashed curve: form factor for the state with $n=1,\kappa=0$, but for the same binding energy (and rms radius) as for the curve 
({\it i}).}\label{ff1}
\end{figure}

The electromagnetic form factor for the state with $n=1,\kappa=0$ (No.1 from the Table \ref{tab1}) is shown in the left panel of  Figure \ref{ff1}.
The electromagnetic form factor for the state with $n=1,\kappa=2$ (No.3 from the Table \ref{tab1}) is shown in the right panel of  Figure \ref{ff1}.
We see that form factor for the abnormal state (right panel, solid curve) \vs \, $Q^2$ decreases much faster than for the normal one (left panel). For $Q^2=1$ 
the form factor for abnormal state is almost $10^3$ times smaller than for the normal one at $Q^2=10$. This fast decrease can be caused either by many-body nature of the abnormal state, or by its small binding energy (by its large size). According to \cite{MMT}, the elastic form factor of a $n$-body system should decrease as $1/(Q^2)^{n-1}$. In Wick-Cutkosky model one can expect even more fast decrease \cite{EPJC}: $\sim 1/(Q^2)^n$. To make choice between these two reasons (many-body nature or small binding energy),  in Figure \ref{ff1} we show by dashed curve the form factor for the normal state but calculated for the same binding energy (and rms radius)  as for abnormal one.
We see that it also quickly decreases, but at $Q^2=1$ it remains ten times larger than the abnormal state form factor. Therefore, fast decrease of the abnormal form factor can be explained by the small binding energy only partially. It indicates another reason which for the abnormal state is its many-body structure. This is in accordance with its very small two-body sector  contribution (see last column in the Table \ref{tab1} for $\kappa=2,4$).

\section{Transition form factors}\label{transit}
\begin{figure}[!h]
\centering
\epsfxsize=5.cm\epsfysize=4.cm\epsfbox{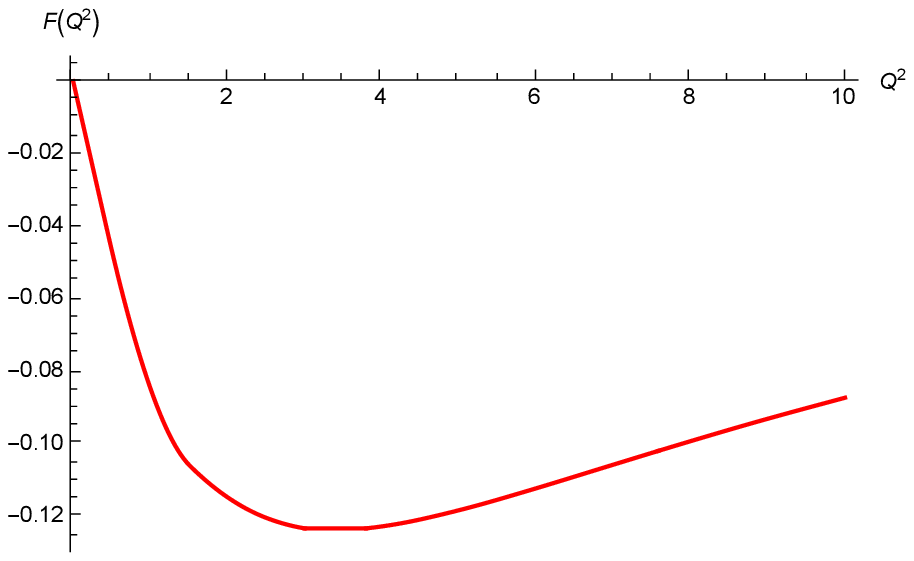}\hspace{1cm}
\epsfxsize=5.cm\epsfysize=4cm\epsfbox{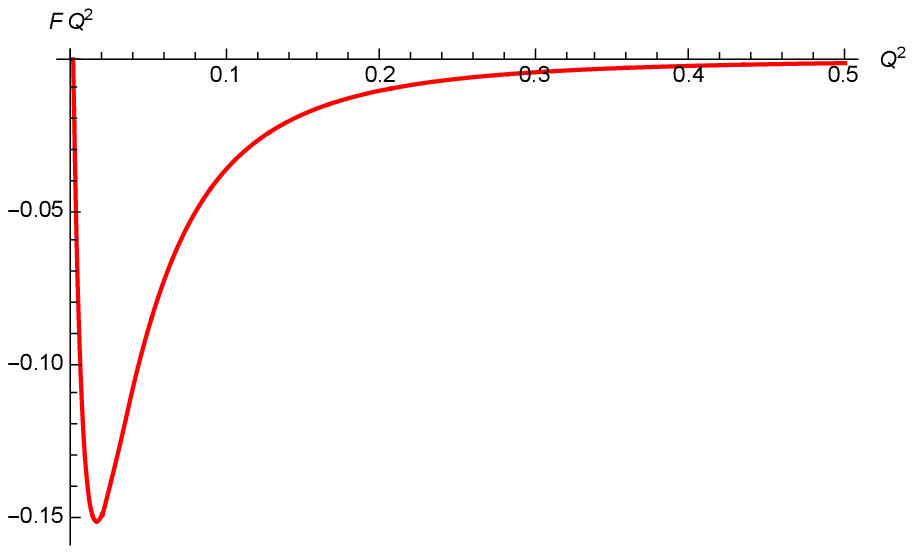} \\
\epsfxsize=5.cm\epsfysize=4.cm\epsfbox{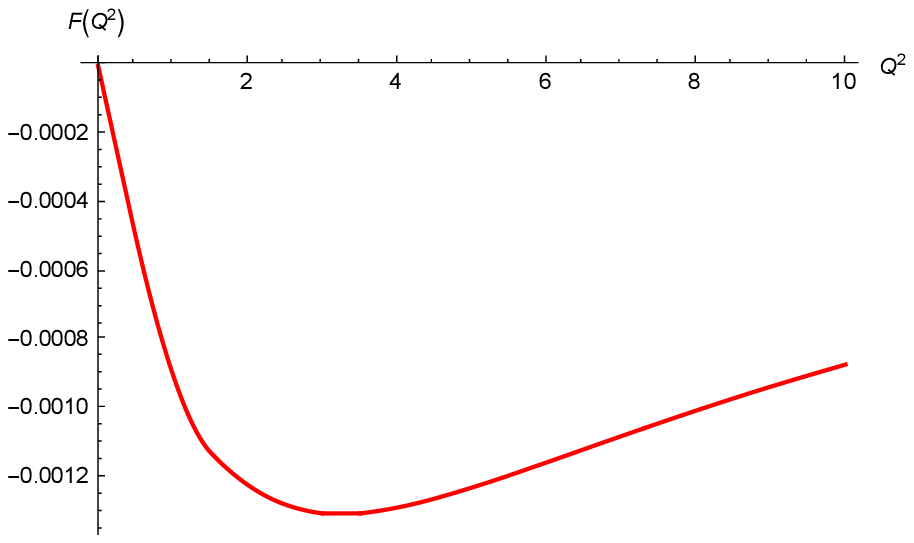}
\caption{Upper left panel: transition form factor between two normal states:\\ 
No. 1 ($n=1,\kappa=0$) $\to$ No. 2 ($n=2, \kappa=0$) from the Table \ref{tab1}.\\
Upper right panel: transition form factor between two abnormal states:\\
No. 3 ($n=1,\kappa=2$) $\to$ No. 4 ($n=2, \kappa=2$).\\
Lower panel: transition form factor between the normal and abnormal states:\\ 
No. 1 ($n=1,\kappa=0$) $\to$ No. 3 ($n=1, \kappa=2$).}
\label{Ftr}
\end{figure}

The drastically different structure of the normal and abnormal states is also confirmed by the transition form factors shown in Figure \ref{Ftr}. At upper part of this Figure the form factors for transition between the states of the same nature are shown: normal $\to$ normal (left panel) and abnormal $\to$ abnormal (right panel). Their maximal absolute values are approximately the same, though the latter varies much more quickly. Whereas, the form factor for the transition normal $\to$ abnormal state, shown in the lower panel, is suppressed by the factor 100. 


\section{Conclusion}\label{concl}
Wick \cite{Wick} and Cutkosky \cite{Cutkosky} have shown that in the Coulomb relativistic problem (two particles interacting by massless exchange,
 analyzed in the framework of the Bethe-Salpeter equation \cite{bs}) the spectrum is not reduced to the shifted Coulomb levels $E_n\propto 1/n^2$.
 For the coupling constant $\alpha>\pi/4$, extra infinite series of levels appear. They are enumerated, for given $n$, by the extra quantum number $\kappa=0,2,4,\ldots$, see (\ref{abnspectr}). For $\kappa=0$ these levels are the usual Coulomb levels $\propto 1/n^2$ (though, beyond  the order $\alpha^2$, 
with the coefficient corrected by the relativistic effects,   see  (\ref{Enrel})). For $\kappa=2,4,\ldots$, they are forming new series, in addition to the Balmer series. They are not given by the Schr\"odinger equation and they disappear in the non-relativistic limit. We have analyzed \cite{EPJC} the Fock sector content of these states. It turned out that, in contrast to the normal states, these extra (abnormal) states are dominated by the exchange massless particles, moving with speed of light. That's why they are absent in the Schr\"odinger equation framework.
 These abnormal states appear as a highly excited levels  in the enough strong field with $\alpha >\pi/4$ that corresponds to the charge $Z>107$.
Their experimental discovery (by analyzing spectrum and elastic and transition form factors) seems not easy but solvable problem. 
 
So far we discussed the systems bound by the electromagnetic interaction.
However, one should also mention glueballs and the hybrid states predicted in QCD.
In principle, the   glueballs originate due to different reason - self-interactions of gluons. They do not contain the constituent quarks. 
Besides, we considered scalar massless colorless exchanges, when the problem of the color compositeness of the hybrid states does not appear. Whereas, the glueballs made of the colored gluons must be colorless. This imposes restrictions of their compositeness. In our opinion,  these differences, however, can be considered as secondary. Generally, the main reason of origin of systems dominated by massless particles is
possibility of easy virtual creation of the latter ones in the intermediate states, independently of particular mechanism of their creation: either self-interaction, or exchanges between constituents, in the ladder approximation, or beyond it. From this general point of view, the states discussed in the present article and the glueballs can be considered as systems of similar nature.

\end{document}